\documentstyle{elsartapp}
\begin{document}
\bibliographystyle{unsrt}
\parindent=1cm

\begin{frontmatter}

\title{Detecting Neutrinos from AGN:\\New Fluxes and Cross Sections}

\author{Gary C. Hill}
\address{Department of Physics and Mathematical Physics\\
           University of Adelaide\\
           Adelaide 5005\\ 
           South Australia\\
    e-mail: ghill@physics.adelaide.edu.au\\}


\begin{abstract}
New information on the structure of the nucleon from the HERA $ep$
collider leads to higher neutrino cross sections for the processes
$\nu_{\mu}(\bar{\nu}_\mu) + N \rightarrow \mu^-(\mu^+) +
 X $
needed to calculate the expected rates of astrophysical neutrino induced
muons in large detectors either under construction,
 or in the design stage. These higher cross sections lead to higher muon
rates for arrival angles where neutrino attenuation in the earth is
less important. On the other hand, new estimates of AGN neutrino fluxes suggest that
the expected muon rates in these detectors may be much lower than previously
calculated. I use the new cross sections to calculate the expected muon 
rates and angular distributions in large detectors
 for a variety of AGN models and compare these rates
with the atmospheric neutrino backrounds
 (from both conventional decay channels and the 
\lq prompt' charmed meson decay channels). If the lowest flux estimates are
correct, there may be diffculties
in  determining the origin
 of a small excess of
muons, due to the large uncertainty
in the rate of
the charmed meson neutrino production channel. However, the more optimistic
AGN neutrino fluxes  should be detectable in
proposed detectors, such as DUMAND-II, AMANDA, NESTOR and Lake Baikal.

\end{abstract}

\end{frontmatter}

\newpage

\section{Introduction}
In this paper I consider the problem of detecting  neutrinos 
from active galactic nuclei (AGN).  Recent predictions
 (see for example \cite{SDSS91,SDSSH92,SS95,SPH92,SP94}) of the
diffuse background of neutrinos from the sum of all AGN are high
enough that water and ice Cerenkov detectors under construction
such as DUMAND \cite{Roberts92}, AMANDA \cite{AMANDA95},
 Lake Baikal \cite{Baikal95} and NESTOR \cite{NESTOR95} should be sensitive
to  muons from the interactions of the neutrinos
 in the earth
 \cite{StanevH92,Stanev94,Stenger91,ZHV92,GHS95,FMRradio,GQRS95}.
 Most recently, Protheroe \cite{P96} has used a new model to calculate 
the rate of gamma ray and neutrino production
in AGN jets. The resulting diffuse neutrino background varies over 
many orders of magnitude depending upon assumptions made about the
relative contributions of various production mechanisms in the AGN system.
 The minimum estimated flux may be beyond the reach of the current 
generation of neutrino detectors mentioned above. 
I have reassessed the detectability of a wide range of 
flux predictions using calculations of the neutrino
cross sections based on 
 nucleon structure function parameterizations \cite{MRSG,FMR95}
(parton distributions)
derived from 
 recent accelerator
experiments \cite{H192,ZEUS92,H193,ZEUS93}.
Use of the new parton distributions leads to higher cross sections and a
corresponding increased probability of a detectable muon being produced in 
the rock surrounding a detector. 
 However, increased neutrino absorption in the earth is an 
important effect and  offsets some of the gains for the longest neutrino
paths through the earth. 
I present the rates and angular 
distributions of muons  for  different AGN spectra  and for the 
flux of atmospheric neutrinos \cite {Lipari93} produced when
 cosmic rays  interact with the earth's atmosphere, including neutrinos
from charmed meson decay \cite{ZHV92,GHS95,TIG95}.

The outline of the paper is as follows.  A brief review
of the   
quark parton model and how it is used to calculate neutrino
 cross sections using accelerator structure function data (parton 
distributions) is presented. 
   I then examine two of the most recent sets of parton
distributions and look at the implications for the total cross sections.    
Finally, I discuss the use of these 
cross sections to estimate the rates and angular
distributions of neutrino
induced muons in large detectors, using a Monte Carlo simulation of the
passage of neutrinos through the earth. This has the advantage of allowing
 a precise treatment of the effects of neutral current scattering, which
reduces the neutrino energies before they finally interact to produce muons.
The prospects for detection of the AGN fluxes above the atmospheric
backgrounds are then examined.

Upon completion of these calculations, I learned of similar work by
Gandhi, Quigg, Reno and Sarcevic \cite{GQRS95}, using another recent
set of parton distributions \cite{CTEQ}, which yield muon rates
consistent with the results presented in this work.

\newpage

\section{Quark-parton model and the neutrino cross sections}

\subsection{Parton model description of charged
 and neutral current neutrino cross sections}

The interaction of a neutrino with a nucleon may be described
 \cite{TDLee,AH89}
in terms of the quark-parton model. 
The nucleon is considered
to be a collection of quasi-free \lq \lq partons'' (quarks and gluons) 
which share the 
nucleon momentum.
A proton consists of three valence quarks  (up, up and down) surrounded 
by a sea of quark/antiquark pairs of all flavours.
The description of a weak interaction of a neutrino (energy $E_\nu$)
with a 
nucleon (mass $M_N$) in the  quark-parton model is described in terms of an
interaction of the neutrino with any one of the individual quarks (which carries
a fraction $x$ of the nucleon momentum).
In a charged current process a $W$ boson is exchanged from a  
muon-neutrino to the quark with the neutrino turning into
a muon (energy $E_\mu$).
 A neutral current process involves the exchange of a
$Z$ boson, with the neutrino remaining a neutrino.
 The square of the 4-momentum transfer is denoted $Q^2$ and the 
variable $y = \nu /E_\nu$ ($\nu = E_\nu - E_\mu$) describes the fractional 
energy transfer.
The cross section for the inclusive processes
$\nu_{\mu}(\bar{\nu}_\mu) +
 N \rightarrow \mu^-(\mu^+) + X$
 and $\nu_\mu(\bar{\nu}_\mu) + N \rightarrow
\nu_\mu(\bar{\nu}_\mu) + X $ can be written in terms of the variables
 $ x $  and 
$ 
                 y = \nu/E_{\nu}
$
(referred to as Bjorken scaling variables)
where $x$,$y$ and $Q^{2}$ are related to the square of the 
centre of mass energy $s = 
  2M_{N}E_{\nu}$, by
        $Q^{2} =  sxy$.

The cross section for the {\em charged current} interaction 
$\nu_{\mu}(\bar{\nu}_\mu) +
 N \rightarrow \mu^-(\mu^+) + X$
 is written in terms of three structure
functions $ F_{1}^{\nu N, \bar{\nu}N}, F_{2}^{\nu N, \bar{\nu}N}$ 
and $F_{3}^{\nu N, \bar{\nu}N} $, but if the transverse 
momentum carried by the quarks in the nucleon is negligible then the
relation $2xF_{1} = F_{2}$ \cite{CG69} reduces this dependence to two structure
functions \cite{GG87}:
\begin{eqnarray}
& & \left(\frac{d^{2}\sigma}{dxdy}\right)^{\nu N, \bar{\nu}N}  
     =  \frac{G_{F}^{2}M_{N}E_{\nu \bar{\nu}}}{\pi} 
    \left[ \frac{M_{W}^{2}}{Q^{2} + M_{W}^{2}}\right]^{2} \nonumber \\ 
& & \;\;\;\;\;\;\;\;\;\;\;\;\;\; \times 
\left\{(1-y+\frac{y^{2}}{2}) F^{\nu N, \bar{\nu}N}_{2}(x,Q^{2}) \pm 
    (y-\frac{y^{2}}{2})xF^{\nu N, \bar{\nu}N}_{3}(x,Q^{2}) \right\} 
       \label{eq:sigma_cc}
\end{eqnarray}
The $\pm$ sign corresponds to the $\nu / \bar{\nu} $ cross sections.
$ M_W $ is the mass of the \linebreak W-boson and  $ G_F (= 
1.16639 \times 10^{-5} \;\rm{GeV}^{-2}) $ is the Fermi constant \cite{PD94}. 
The cross section in equation \ref{eq:sigma_cc} is expressed in 
 natural units where $\hbar = c = 1$ yielding a result in units of
 GeV$^{-2}$, to convert  
to a cross section in cm$^{2}$ we multiply
 by $0.38939 \times 10^{-27}$ cm$^{2} \; \rm{GeV}^{2}$ \cite{AH89}.
 Equation \ref{eq:sigma_cc} also holds for $\nu_e(\bar{\nu}_e)
\rightarrow e^+(e^-)$.

The structure functions ${F}_{2}$ and ${F}_{3}$ are defined in terms of
  quark distributions \cite{deGroot79}:
\begin{eqnarray}
    {F}^{\nu N, \bar{\nu}N}_{2} = x(q + \bar{q }) \nonumber \\
\end{eqnarray}
and
\begin{eqnarray}
     F_{3}^{\nu N, \bar{\nu}N}  =  q - \bar{q} \pm 2\{[s-c] + [b-t]\}
\end{eqnarray}
where
\begin{eqnarray}
     q & = & u + d + s + c + b + t
      \nonumber \\
   \bar{q} & = & \bar{u} + \bar{d} + \bar{s} +
        \bar{c} + \bar{b}  + \bar{t}
\end{eqnarray}
are the sums of the quark/antiquark distributions ($u,d,s,$ etc). Each
quark distribution
  is a function
of $x$ and $Q^2$  describing the  number of quarks
of that flavour having a fractional momentum in the range $x \rightarrow
x + dx$, at momentum scale $Q^2$. Note  that the $\pm$ sign again corresponds
     to the $\nu / \bar{\nu} $ cross sections.


The {\em neutral current} cross sections
$\nu_\mu(\bar{\nu}_\mu) + N \rightarrow
\nu_\mu(\bar{\nu}_\mu) + X $ are expressed in the same form as 
equation \ref{eq:sigma_cc},
 but with $M_W$ replaced by $M_Z$ and the structure
functions $F_2$ and $F_3$ written as \cite{Buras80}
\begin{eqnarray}
F_2 & = & x(q+\bar{q})
[\delta^2_1 + \delta^2_2 + \delta^2_3 + \delta^2_4 ] \nonumber \\
   & + & 2x[(c-s)+(t-b)][\delta^2_1 + \delta^2_3 - \delta^2_2 - \delta^2_4 ]
\end{eqnarray}
and
\begin{equation}
F_3 = (q-\bar{q})[\delta^2_1 + \delta^2_2 - \delta^2_3 - \delta^2_4 ]
\end{equation}
The $\delta$  factors are the chiral couplings, defined in terms
of the weak mixing parameter $ x_W = \sin ^2 \theta = 0.225$ \cite{PD94}

\begin{eqnarray}
      \delta_1  & = & \frac{1}{2} - \frac{2}{3}x_W \nonumber \\
      \delta_2  & = & -\frac{1}{2} + \frac{1}{3}x_W \nonumber \\
      \delta_3  & = & - \frac{2}{3}x_W \nonumber \\
      \delta_4  & = & \frac{1}{3}x_W
\end{eqnarray}

\subsection{Parton distributions derived from recent accelerator data}
\label{partons}

For this work the cross sections have been calculated using sets of
parton distributions \cite{MRSG,FMR95}
 fitted to the latest HERA deep inelastic
scattering  data \cite{H192,ZEUS92,H193,ZEUS93}. The H1 and ZEUS
collaborations have extracted the electromagnetic structure function 
$F^{ep}_{2}$ from  collisions of  26.7 GeV electrons with 820 GeV 
 protons. $F^{ep}_{2}$ is the sum of the individual parton distributions
weighted by the squares of their electric charges :
\begin{equation}
F^{ep}_{2} = \frac{4}{9}(u+\bar{u}) +  \frac{1}{9}(d+\bar{d})
       + \frac{4}{9}(c+\bar{c}) + \frac{1}{9}(s+\bar{s})+
      \frac{1}{9}(b+\bar{b})
\end{equation}
From the $F^{ep}_{2}$ measurement the individual quark 
distributions can be calculated and used to form the structure
functions $F_{2}^{\nu N} $ and $F_3 $ needed for the 
neutrino nucleon cross section calculation.
However, to calculate the cross section in the energy range $10^5-10^8$
GeV, we need to know the parton distributions at higher $Q^2$ 
than the HERA measurements. QCD theory 
\cite{GL72,GL72b,AP77,Dok77} describes
the evolution of the parton distributions to higher $Q^2$.
I consider two different solutions of the evolution equations, which
predict sufficiently different behaviour in the unknown regions of
$Q^2$ and $x$ to give neutrino cross sections up to a factor of two
bigger than previous calculations \cite{RQ88,QRW86,RenoH92} in the energy range of interest.
Martin, Roberts and Stirling have presented several sets of 
 parton distributions \cite{MRS93,MRS93b,MRS94}
 based on the increasing amount of accelerator data.
Their latest parton set (\lq \lq MRS G") \cite{MRSG} is a parameterisation that
 includes the recent HERA $F_2$ data. The  MRS fits supply 
individual parton distributions $ u_{v},d_{v},
u_{s}, d_{s}, s_{s}, c_{s}, b_{s} $ which can be used to  form
$F_2$ and $F_3$.

In their recent work \cite{FMR95} Frichter, McKay and Ralston have effectively
 fitted a
 two parameter ($A,\mu$) solution of the structure function evolution
 equations
 to the HERA $ F^{ep}_{2} $ data :
\begin{eqnarray} \label{FMRF2ep}
F^{ep}_{2}(x,Q^{2}) & = & A x^{-\mu}
   \frac{P_{qg}(1+\mu,Q^{2})}{P_{gg}(1+\mu,Q^{2})} \nonumber \\ 
& \times &
 \exp \left\{ \ln\ln \left(\frac{Q^{2}}{\Lambda^{2}} \right) 
   \ln \left( \frac{Q^{2}}{\Lambda^{2}} \right)
       P_{gg}(1+\mu,Q^{2}) \right\}
\end{eqnarray}
 $ \Lambda $ is the QCD scale parameter (taken as $ 200 MeV $
in the FMR result) and $P_{qg}(1+\mu,Q^{2})$ and $P_{gg}(1+\mu,Q^{2})$
are the $1+\mu$ moments of the quark-gluon and gluon-gluon splitting
functions, evaluated at next to leading order \cite{Buras80,FKL81,Gluck90}.
In order to obtain $F^{\nu N}_{2} $,
FMR make the assumption that the quark ($q_i$) and anti-quark ($\bar{q}_{i}$)
distributions of the $ith$ flavour are equal, and that 
\begin{equation}
       q_{i} = \bar{q}_{i} = u = d = s = 2c = 2b
\end{equation}
Substituting into $F^{ep}_{2}$ and $F^{\nu N}_{2} $
yields $F^{ep}_{2} = \frac{17}{9} q_{i}(x,Q^{2}) $ and 
$F^{\nu N}_{2} = 8 q_{i}(x,Q^{2}) $ from which we obtain :
\begin{equation} \label{FMRF2nuen}
F^{\nu N}_{2} = \frac{72}{17} F^{ep}_{2}
\end{equation}
A least squares minimization to the 1992 and 1993 HERA $F_2^{ep}$ data
yields best fit parameters $ A = 0.008829 $ and $\mu = 0.3856$.

\subsection{The total neutrino charged and neutral current cross sections}
 To obtain the total cross sections I have integrated equation 
\ref{eq:sigma_cc} 
 using the MRS G and FMR parton distributions, for both the charged and
neutral current forms of $F_2$ and $F_3$.

  The MRS G distributions cover the full range $ 0 < x < 1 $
 required, but the FMR result applies only at $ x \sim 0.2 $ or less.
 In order to evaluate the FMR cross section  over
the full $ x $ range, I have used the FMR result for $ x < 0.2 $ and the MRS G
result for $ x \geq  0.2 $. 
The MRS G parton distributions are valid down to $x \sim 10^{-5}$, which is
adequate for the calculation of the cross sections in the energy
range considered in this work ($E_\nu < 10^8$ GeV).  (Note that the
double-logarithmic approximation \cite{GLR83} may be used to
 extrapolate the parton 
distributions to lower $x$ for cross section calculations \cite{RQ88,QRW86,RenoH92} where
$E_\nu > 10^8 $GeV.) 

Note also that I have combined a set of {\em next-to-leading order} (NLO)
parton distributions (MRS G) calculated in the $ \overline{MS} $
 renormalization 
convention  with the {\em leading order} (LO) form of the structure 
functions $F_2 $ and $ F_3 $. The  NLO expressions for $F_2 $ and $ F_3 $
(obtained by convolving the quark and gluon distributions with the
Wilson coefficients \cite{Wally95}) will differ from the LO expressions
by terms with coefficients $\alpha_S(Q^2)/2\pi $, where $\alpha_S(Q^2)$ is
the QCD coupling constant, which decreases as $Q^2$ increases. 
In the regions of high
$Q^2$ ($\sim M_W^2$) important for the neutrino energies considered in this
work $\alpha_S(Q^2)/2\pi $ has a value of $\sim 0.02 $ and the resulting 
differences between the LO and NLO structure functions, and hence the difference
in the resulting cross sections, will be small ($\sim$ a few percent).
Given the very large uncertainties in the astrophysical assumptions in this
work, this is an acceptable approximation.

\begin{figure}[htbp]
\vspace{100mm}
\includegraphics{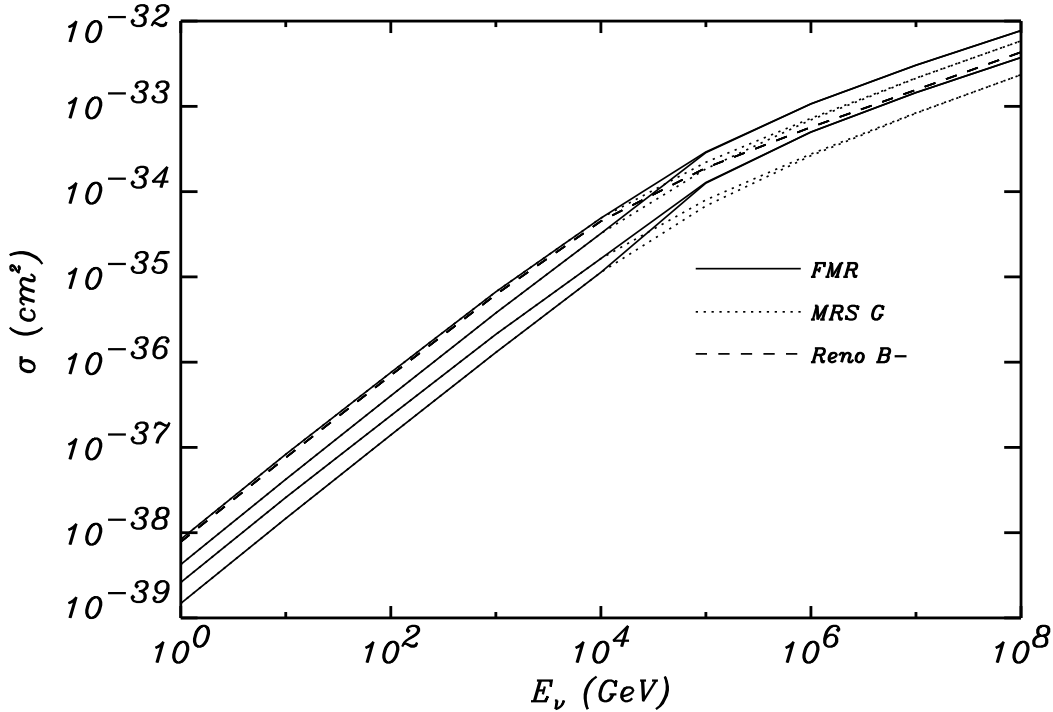}
\caption{\label{crossfinal} Total $\nu$ and $\bar{\nu}$ cross sections 
calculated using recent parton distributions. At low energy the curves
represent (in decreasing order of cross section) $\nu$ - charged current,
$\bar{\nu}$ - charged current,  $\nu$ - neutral current and
$\bar{\nu}$ - neutral current. At high energy, neutrino and anti-neutrino
cross sections are equal.}
\end{figure}

 The total  neutrino/anti-neutrino charged/neutral current cross sections
obtained by using the FMR and MRS G parton distributions are shown in
figure \ref{crossfinal}, along with the \lq \lq $B-$ '' neutrino charged current
 estimate by Reno \cite{RenoH92}.  Above $ 10^5 $ GeV neutrino and
anti-neutrino cross sections are equal and  the FMR cross sections are roughly  1.5 times
bigger than the MRS G predictions and  2 times bigger than the Reno $B-$
prediction.  The recent calculations of Gandhi {\em et al} \cite{GQRS95}, 
using the CTEQ3 parton distributions \cite{CTEQ}, yield a neutrino charged
current cross section lower than using MRS G by about 7\% at 10$^3$ GeV, with
difference rising to 23\% at 10$^8$ GeV. The CTEQ3 cross section is
bigger than the Reno $B-$ cross section by 1-10\% in the same energy range.

 \section{Astrophysical neutrino fluxes }
 
\begin{figure}[htbp]
\vspace{100mm}
\includegraphics{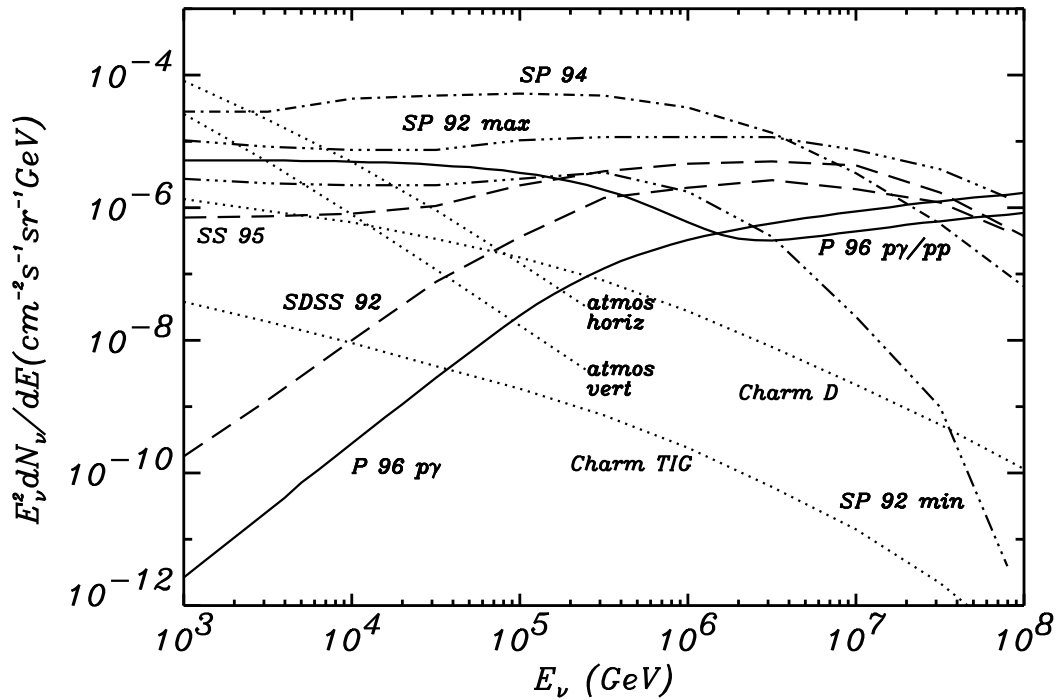}
\caption{\label{agnspec} Diffuse isotropic neutrino fluxes from the sum of
all active galactic 
nuclei predicted by various models. The angle dependent
neutrino flux produced from 
cosmic ray interactions in the earth's atmosphere 
is also shown for horizontal
and vertical directions. The isotropic neutrino flux produced from prompt
decays of charmed mesons in the atmosphere is also shown.}
\end{figure}

 The reported observations \cite{SS83,JLE83,PCG84,Yodh92} of UHE $\gamma$--rays from binary X-ray 
systems in the early 1980s led to predictions of neutrino fluxes
from such objects. The assumption was that protons accelerated in the
vicinity of the compact object would interact with other matter in the
system to give both neutral and charged pions, the neutral pions decaying
to produce the observed $\gamma$--rays and the charged pions producing
a neutrino flux. Calculations were made \cite{GG87,RQ88,GS85,GS85a,KTW85}
of the expected neutrino flux
by scaling the models to the observed $\gamma$--ray flux. However the
non-confirmation of the $\gamma$--ray observations by much larger and
more sensitive air shower arrays such as CASA
 \cite{CASA92,CASA95a,CASA95b,CASA95c,CASA95d} has led to upper limits
on the  $\gamma$--ray fluxes that are orders of magnitude lower than
the original detection claims, dampening expectation of observable
 neutrino fluxes from binary X-ray systems. 

Attention in recent times has focussed on the production of neutrinos
in active galactic nuclei, where particle acceleration may be occuring
at accretion shocks in the vicinity of massive black holes. 
 Stecker, Done, Salamon and Sommers (SDSS) \cite{SDSS91,SDSSH92} calculated  the neutrino flux resulting
from interactions of protons with energy  above the threshold energy for
 photoproduction on UV photons ($p\gamma \rightarrow
\Delta \rightarrow n\pi^+ $).  Szabo and Protheroe (SP) \cite{SPH92,SP94}
 included all the subsequent interactions
of the protons after the initial photoproduction.
Luminosity functions \cite{MD91,MT89} (derived from X-ray and gamma-ray
measurements - describing the
luminosity of AGN as a function of redshift)  and the 
relationship between the neutrino and X-ray and gamma-ray outputs
 in a given AGN model
were used to calculate the neutrino output from the sum of all AGN
in the universe.
     More recently Stecker and Salamon \cite{SS95} revised their original
model and calculated diffuse neutrino backgrounds for quasar cores and
blazar jets.

 Protheroe \cite{P96} has calculated the neutrino and gamma-ray fluxes
from proton acceleration in \lq blobs ' of matter in AGN jets. Two major
production mechanisms were considered. Accelerated protons may leave the
blob and interact with the radiation field of the accretion disk
($p \gamma$ interactions), or interact with each other or with other
matter in the vicinity ($pp$ interactions). The observed time variability
of gamma-rays from AGN such as MKR421 place constraints on the relative
contribution of each mechanism.
The $pp$ interactions have a long characteristic time scale and therefore
 cannot
produce variability of the order of a day. The $p\gamma$ contribution,
however, can vary rapidly and must be responsible for the time variation.
To produce an intensity change of a factor of 2 over a short time scale,
 the $pp$ interaction can contribute no more than about 50\% to the
total output. The $pp$ interactions produce the most neutrinos, so the
maximum allowable neutrino flux comes from equal contributions of $pp$ and
$p\gamma$ and the minimum flux from 100\% $p\gamma$.
 The luminosity function \cite{EGRET} derived from the
EGRET survey and the results from the SAS-2 experiment \cite{SAS2} have
been used to produce a diffuse neutrino background estimate.

Figure  \ref{agnspec} shows the resulting neutrino fluxes from a number
of different models. Note that the fluxes are multiplied by $E^2$ to
reduce the range of the plot. Curves labelled \lq SP' (dot-dashed lines)
are predictions
from the earlier Szabo and Protheroe models. \lq SDSS92 ' is the
Stecker {\em et al} $p\gamma$ flux, while \lq SS95' is the sum of the
quasar core and blazar jet contributions in the Stecker and Salamon model 
(dashed lines). The new Protheroe model is shown in solid lines for the
limiting cases, the
 minimum expectation coming from $p\gamma$ interactions
only and the maximum from equal contributions from $pp$ and $p\gamma$ 
interactions. 

The major backgrounds to searches
 for AGN fluxes come from the decay
of kaons, pions  and muons produced from the interactions of cosmic rays in the
earth's atmosphere. These fluxes show an angular dependence; in the
horizontal direction the fluxes at the surface are greatest as the kaons, pions  and muons
traverse more atmosphere and have more time to decay. The results of Lipari
\cite{Lipari93} are shown. Another potential background comes from the 
prompt decay of charmed particles produced in cosmic ray interactions in the
atmosphere. The expected neutrino fluxes should have a flatter energy spectrum
and not show an angular dependence as the particles mostly decay before
they can interact.   
Zas, Halzen and V\'{a}zquez \cite{ZHV92} have calculated the expected 
neutrino fluxes from charmed meson decay under various assumptions of
the charm production cross section and have computed muon rates in 
downward looking detectors.  Gaisser, Halzen and Stanev \cite{GHS95}
conclude  that  model  B is an upper limit to the neutrino fluxes in
the region $E_{\nu} > 100$ TeV. However, model D predicts higher fluxes
at lower neutrino energies and gives overall  the highest muon rates in
downward looking detectors of the two cases. I therefore use model D to
predict the upper limit on the muon rate from charm production.
As a lower limit, I take the prediction of Thunman, Ingelman and Gondolo
\cite{TIG95} which gives neutrino fluxes nearly 100 times smaller than
the ZHV model D. It is clear that there is a large uncertainty in the
potential contribution of charmed meson decay to the neutrino induced
muon fluxes. The implications of this on the
detectability of the AGN fluxes will be addressed in the next section.
\newpage

\section{Neutrino induced muon rates in large detectors}

\subsection{The neutrino interaction Monte Carlo}
Cerenkov detectors (DUMAND, AMANDA, NESTOR, Lake Baikal) 
detect upward going muons from charged current interactions of
neutrinos in the earth. The muons can be produced many kilometres 
from the detector and still survive to be detected via their Cerenkov
emissions in water or ice. 
Calculations of the expected muon fluxes from AGN
 \cite{StanevH92,Stanev94,Stenger91,ZHV92,GHS95} have been made using various
parton distributions and muon energy loss assumptions. 
 In their original cross section paper \cite{FMR95},
 Frichter, McKay and Ralston calculated the
rates of charged lepton production in a cubic kilometre volume of ice and
noted that despite the increased attenuation with their new cross section,
the increased interaction probability more than compensated for this, resulting
in increased lepton production. In later work \cite{FMRradio},
 they calculated the
specific rate expectations for a radio receiver in Antarctic ice, detecting
radio Cerenkov radiation from electrons. The highest of the Szabo and 
Protheroe fluxes \cite{SP94} and the Stecker {\em et al} flux \cite{SDSS91}
were used, the results suggesting these fluxes would be readily 
detectable with a single antenna buried at a depth of 600 metres. 
 Zas, Halzen and V\'{a}zquez \cite{ZHV92} calculated the detection rates
of neutrino induced horizontal air showers and underground muons,
again for SDSS and SP fluxes. These calculations included the expected
rates from charm induced neutrinos. The cross sections were calculated 
with the KMRSB-5 \cite{KMRS} parton distributions which give total
cross sections of about the size of the Reno $B-$ estimate.
 Most recently, Gandhi, Quigg, Reno and Sarcevic  \cite{GQRS95} (GQRS),
have evaluated the cross sections for a variety of parton distributions
and have then used the CTEQ3 parton distributions \cite{CTEQ} to estimate
muon rates from astrophysical sources. 
Due to the relative differences in the
final cross sections,
 the CTEQ3 partons will yield muon rates closer to those
using Reno $B-$ than  MRS G. 

The calculation of  muon fluxes induced by neutrino interactions in the
earth is complicated by  neutrinos  undergoing neutral 
current interactions in the earth (which reduce their energies but do not
remove them from the beam)
before they finally interact to 
produce a muon. The final number of neutrinos in a given energy bin
after passage through the earth is reduced by  both
charged and neutral current interactions, but some of the total
loss is offset by originally higher energy neutrinos having moved
down in energy
due to neutral current interactions. 

 An effective total cross section can be calculated,
which is dependent on the original source spectrum. This approach was 
followed by Berezinsky {\em et al} \cite{BGZR86} 
where the total cross section was taken as the sum of neutral and 
charged current cross sections, with a correction term included to 
account for the source dependent
neutral current neutrino regeneration effect.
 
As an alternative approach to the neutral current effect,
 I have
developed
a Monte Carlo simulation \cite{MyPhD}
to model the passage of the neutrinos through the
earth, taking into account the relative likelihood of neutral and
charged current interactions, and selecting the final neutrino or muon
energy from the energy dependent $y$ distributions. A simple weighting 
system was devised (and generalized to account for neutrinos that undergo
none, one, two or more neutral current scatters before a charged current 
interaction) to keep the computing time reasonable. Neutrinos and
anti-neutrinos were treated separately using the appropriate cross
sections.
 The column density of nucleons to be traversed by a
neutrino as a function of nadir angle is calculated using the 
Preliminary Reference Earth Model \cite{PREM}, which describes
 how the density of
the earth varies with distance from the centre.
The final energy of a muon reaching the detector is determined using the 
muon propagation code of Lipari and Stanev \cite{SL91}. The muon energy losses due to
bremsstrahlung \cite{PS68}, $e^{+}e^{-}$ pair production
\cite{KP71} and photonuclear 
interactions \cite{Bezru81} are treated stochastically in order to give the correct
final fluctuations in muon ranges. Ionization losses are also
taken into account. Only muons which are produced within the final
tens of kilometres from the detector will actually survive to the 
detector, so for these calculations I use \lq \lq standard rock'' ($ Z = 11,
A = 22, \rho = 2.65 $ g cm$^{-3}$) for the propagation.

\subsection{Integral neutrino induced muon rates}
I have calculated the expected muon rates using the cross sections Reno $B-$,
 FMR and MRS G resulting
from the parton distributions described in section \ref{partons} which
are derived from the latest accelerator structure function data.
 The higher cross sections result in greater attenuation of neutrinos
within the earth, but this effect is offset by a greater probability
of muon production by the surviving neutrinos
in the final kilometres of rock before a detector,
leading to greater expected muon rates.
It is  important to remember that the effective
area of a detector will have an energy and arrival angle
dependence and that the overall
detection rate will depend upon the energy and angular spectrum of muons
incident upon the detector. For example, the effective area for muon
detection (averaged over arrival direction) \cite{Stenger92} by the
DUMAND-II detector ($20000$m$^2$ geometric area) varies from 
about $5000$m$^2$ at 10 GeV (the array does not trigger on all muons)
up to $20000$m$^2$ at 5 TeV and goes up to $60000$m$^2$  
at $10^7$ GeV (the array triggers on muons that pass outside the
detector volume). The muon rates calculated here assuming a spherical
detector with a constant 20000m$^2$ effective area for all arrival
directions and all muon energies 
 will therefore be approximate
lower bounds on the numbers of muons detected for muons above 
about 1 TeV (near the energy at which the effective area equals the
geometric area). Due to the increase in effective area at higher 
energies, the muon rates for flatter spectra will be enhanced over
those of steeper spectra compared to the numbers presented here which
are calculated for an energy independent effective area.
 Keeping this in mind, we examine the 
predicted muon rates induced by AGN neutrinos.

\begin{figure}[htbp]
\vspace{100mm}
\includegraphics{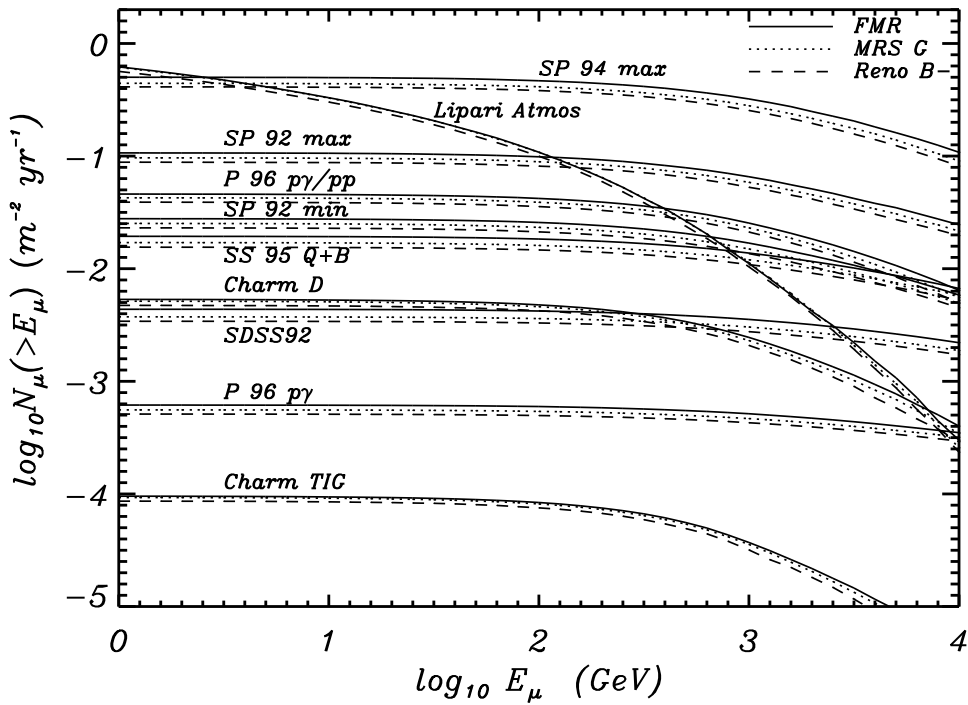}
\caption{\label{intmuonrates} Integral neutrino-induced muon rates in
downward looking detectors as a function of neutrino source spectrum
and parton distribution used in the 
 cross section estimate.}
\end{figure}

 The integral muon rates expected
in a downward looking detector ($2\pi {\rm sr} $ field of view) are shown in
figure \ref{intmuonrates} for the three different cross sections and nine 
different neutrino spectra of figure \ref{agnspec}. 
 The predicted muon fluxes span several orders of magnitude. The SP94
flux exceeds the atmospheric background for a muon energy cut of a 
few GeV. At the other extreme, the P96p$\gamma$ does not exceed the 
atmospheric background until an energy of around 10 TeV.
The wide variation in the expected 
number of muons from charm induced neutrinos
is evident. The TIG neutrino flux induces a very low rate of muons, 
around 1/10 of the lowest AGN flux (P96p$\gamma$), and should not pose
a background problem at any muon energy cut. However, the ZHV D neutrino
flux results in a muon rate exceeding that of the SDDS92 flux for muon
energies up to $\sim$400 GeV, and roughly equal to that of the conventional
atmospheric and P96p$\gamma$ fluxes at a muon energy cut of 10 TeV.
This great uncertainty suggests
that there would be  difficulties in identifying
 the origins of any observed muons should the lowest AGN flux be
correct.  

The results presented here are not directly comparable to those of
Gandhi {\em et al} \cite{GQRS95},
 as the neutrino source spectra used in both cases
are not identical. GQRS made a broken power law approximation to the
SP94 (b=1) neutrino flux; this yields muon rates less than the SP94
curve in figure \ref{intmuonrates}, which comes from using the highest
of the SP94 fluxes.  GQRS also calculated muon rates for the SS95 neutrino
flux, but only for the quasar core component, whereas I have used the
sum of quasar core and blazar jet components. Also, GQRS treat neutral
currents by taking two cross sections: the first the sum of charged and
neutral current cross sections, which gives a lower limit to the
muon rate; the second the charged current cross section alone,
which will give an upper limit to the muon rate.
  To check the consistency of the two sets of results, I have used the
forms of the neutrino fluxes as implemented by GQRS with the 
MRS G and Reno $B-$ cross sections and find that the
muon rates obtained are consistent with the differences  in
 cross sections  in the two sets of work.

\subsection{Limits from the Fr\'{e}jus experiment}
The SP94 and SP92max fluxes have already been excluded by an analysis
of the energy distribution of horizontal neutrino-induced muons 
in the Fr\'{e}jus experiment \cite{Rhode94,Rhode95}.
The Fr\'{e}jus detector
was a fine grain tracking calorimeter operated  deep underground, with
the ability to accurately determine the energy of the through going
muons from the energy deposited from electromagnetic showering.
At near horizontal angles the background atmospheric muon flux is
sufficiently suppressed by the rock overburden to allow a measurement
of the rate of atmospheric and AGN induced muons.
The number of muons (producing an observed energy loss in the 
detector greater than the maximum observed energy loss)
 expected from each AGN spectrum has been calculated
\cite{GHS95,Rhode94,Rhode95,Gaisser93}
taking into account the rock thickness, muon propagation and the 
increasing efficiency of the detector with increasing muon energy. 
The high fluxes SP94 and SP92max predict
more muons than the
 90\% confidence level upper limit of 2.3 muons and are therefore
excluded by this experiment. All the other fluxes considered predict
less than 2.3 muons and are therefore not excluded  
by  the Fr\'{e}jus  result.

\subsection{Effects of the higher neutrino cross sections}
\begin{figure}[htbp]
\vspace{200mm}
\includegraphics{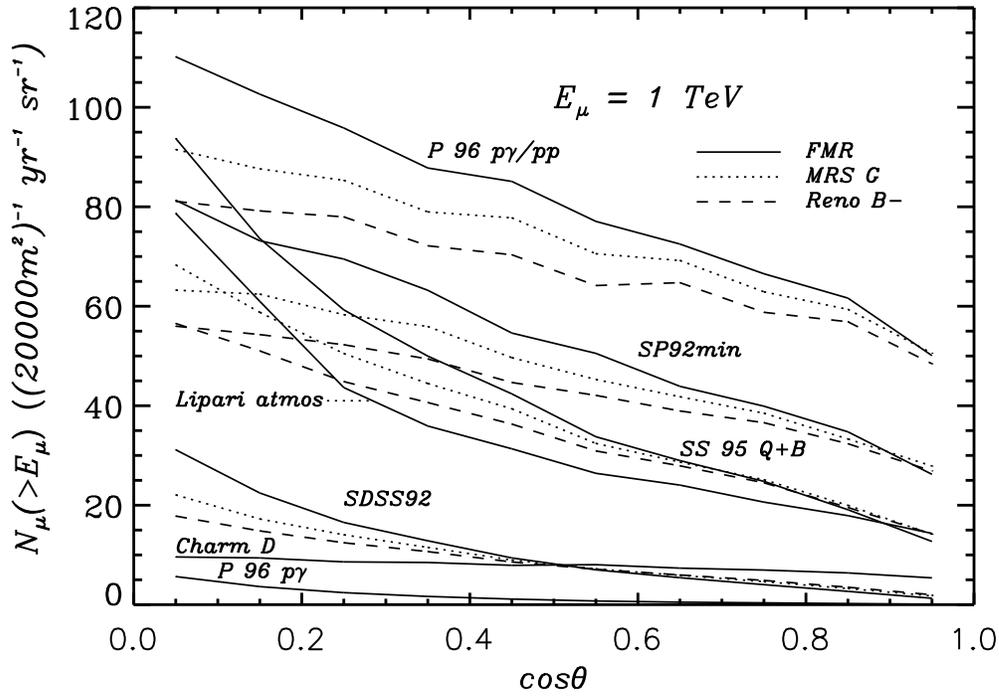}
\includegraphics{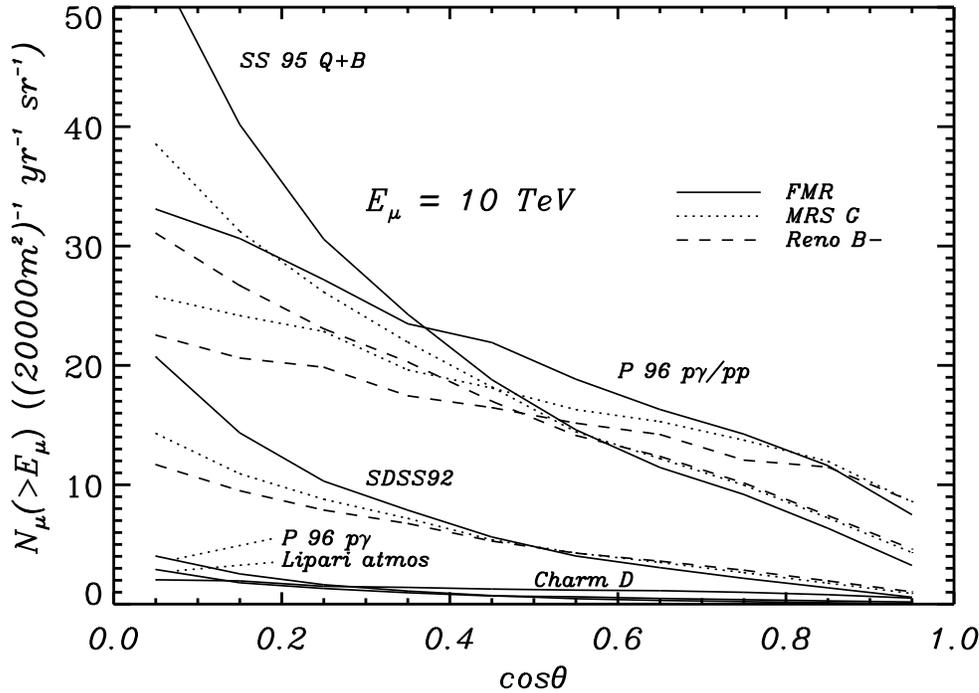}

\caption{\label{muonang103} Muon angular distributions for a
$20000$m$^2$ downward
looking detector employing a
 muon energy cuts of $10^3$ and $10^4$ GeV, for neutrino fluxes not excluded by
 the Fr\'{e}jus experiment. For clarity, only the FMR cross section results
are shown for the smaller fluxes.}
\end{figure}

The angular distributions of the
muon fluxes for 
muon energies above $10^3$ GeV and $10^4$
 GeV are shown in figure \ref{muonang103} for 
a collecting area of 20000 m$^2$, which corresponds to that of the
proposed nine string DUMAND-II array.
The increases in the overall muon rates predicted
by the FMR cross section are not uniform over arrival direction.
It is clear that most of the increases come from the highest nadir angles
 (muons coming from angles 0 to $\sim$30 degrees down from the
horizon) where the thickness of the earth traversed by the
neutrinos is smaller and attenuation less important while the likelihood
of muon production near the detector is greater.
The rate of directly upgoing muons is very
similar for all three cross sections considered - the increased rate
of muon production in the final kilometres of rock is balanced by
increased neutrino attenuation in the earth.
 The increase in muon rates comes from arrival directions where
surface or shallow detectors are most susceptible to noise contamination
from downward going atmospheric muons mis-reconstructed as horizontal
or upward going. However, deep underwater detectors which are better
shielded
from atmospheric muons will be sensitive to these increased near
horizontal fluxes. For example, the DUMAND-II detector  is expected to 
be sensitive to
muons from downgoing neutrinos up to $\sim$20 degrees above the horizon.

\begin{table}[htbp]
\caption{\label{ratetable}Integral muon rates in a $20000$m$^2$
downward looking detector for nadir angles $0-60^o$ (most upgoing) and
 $60-90^o$ (nearer horizontal) for various neutrino spectra,
 cross sections and minimum muon energies.
 }
\vspace{0.5cm}
\begin{tabular}{|c|c|c|c|c|c|c|c|} \hline
    \multicolumn{2}{|c|}{Muon energy cut $\rightarrow$}
   & \multicolumn{2}{c|}{$E_{\mu} > 10^2$ GeV} &
      \multicolumn{2}{c|}{$E_{\mu} > 10^3$ GeV}  &
      \multicolumn{2}{c|}{$E_{\mu} > 10^4$ GeV}  \\  \hline
\multicolumn{2}{|c|}{Nadir angle $\rightarrow$}
    & $\theta>60^\circ $ & $\theta<60^\circ $
&$\theta>60^\circ $   &$\theta<60^\circ $  &$\theta>60^\circ $  &
$\theta<60^\circ $   \\
    \hline
     & Atmos & 1421 & 744 & 157 & 65 & 4.8 & 1.2 \\
     & SDSS  &  68  &  16 &  58 & 13 & 37 & 7.0 \\
     & SP92min & 315 & 200 & 215 & 123 & 83 & 34 \\
FMR  & SS 95 Q+B & 258  & 109  &  200 & 75 & 105 & 28  \\
     & P 96 p$\gamma$/pp & 483 & 358 & 303 & 206 & 86 & 43 \\
     & P 96 p$\gamma$ & 10.5 & 1.5 & 9.1 & 1.2 & 6.3 & 0.7 \\
     & Charm D&52  & 43& 28 &21 &5.1 & 2.9\\
     & Charm TIG&0.9  & 0.8& 0.4 &0.3 &0.06 & 0.03\\
       \hline
     & Atmos & 1394 & 767 & 158 & 67 & 4.1 & 1.1 \\
     & SDSS & 55 & 17 & 46 & 14 & 29 & 8.2 \\
     & SP92min & 272 & 193 & 182 & 117 & 65 & 33 \\
MRS G & SS 95 Q+B & 214  & 108  & 164   & 76  & 85 & 30 \\
     & P 96 p$\gamma$/pp & 434 & 343 & 265 & 196 & 69 & 41 \\
     & P 96 p$\gamma$ & 9.4 & 1.8 & 7.9 & 1.5 & 5.5 & 0.9 \\
     & Charm D&49  & 43& 25 &21 &4.2 & 2.7\\
     & Charm TIG&0.9  & 0.8& 0.4 &0.3 &0.05 & 0.03\\
       \hline

     & Atmos & 1284 & 710 & 145 & 63 & 3.6 & 1.1 \\
     & SDSS & 48 & 18 & 40 & 15 & 26 & 8.6 \\
     & SP92min & 244 & 181 & 161 & 111 & 58 & 32 \\
Reno B- & SS 95 Q+B & 189   & 105   & 144   & 74 &  74 & 31  \\
     & P 96 p$\gamma$/pp & 393 & 320 & 239 & 184 & 61 & 39 \\
     & P 96 p$\gamma$ & 8.0 & 1.9 & 7.0 & 1.6 & 4.9 & 1.0 \\
     & Charm D&46  & 39& 24 &18 &3.4 & 2.4\\
     & Charm TIG&0.8  & 0.7& 0.4 &0.3 &0.04 & 0.03\\
       \hline
\end{tabular}
\vspace{0.5cm}
\end{table}

\subsection{Detectability of predicted AGN fluxes} 
 Here we consider only those
neutrino predictions that have not been ruled out by the Fr\'{e}jus
 experiment. The total muons expected for nadir angles greater and less
than 60 degrees as a function of muon energy cut and cross section for
each neutrino source are summarized in table \ref{ratetable}.
It is evident that the highest of these fluxes may be observable
over the atmospheric background for a 100 GeV muon energy cut ($\sim$300-500
events per year above $\sim$2000 background) but that the smaller fluxes
(10-100 events per year) would not.

For a muon energy cut of 1 TeV, the atmospheric background is significantly
reduced and the fluxes P96$p\gamma /pp$, SP92min and SS95 Q+B are now
greater than the background and should be observable. The lower SDSS92
flux would produce $\sim$55-70 events per year above a 225 background and
should be observable over several years but an uncertain
contribution from charm induced neutrinos ($\sim$1-50 muons) would
complicate matters. However figure \ref{muonang103} shows that the
charm induced muons have a flat angular distribution and this might
aid in identification.  The P96$p\gamma$
is well below the background and given the charm uncertainty, would
seem not to be detectable.

Moving to a 10 TeV cut reduces the atmospheric background  to $\sim$5
per year, making all the fluxes readily observable, except for the
lower bound of P96$p\gamma$. This latter flux, the atmospheric background and
the highest charm decay estimate (ZHV D) 
produce roughly equal numbers of muons.
 Given that the charm induced muon rate could
actually be practically zero (as table \ref{ratetable} shows), this
would make a differentiation between charm induced and AGN induced muon 
fluxes difficult, if only a small excess above the
expected conventional atmospheric rate was seen.
 One could not simply do a counting experiment as the excess above the
conventional atmospheric background could come from either of two 
sources. An examination of the angular distributions or differential
energy spectra of the detected muons would be needed.

\begin{figure}[htb]
\vspace{100mm}
\includegraphics{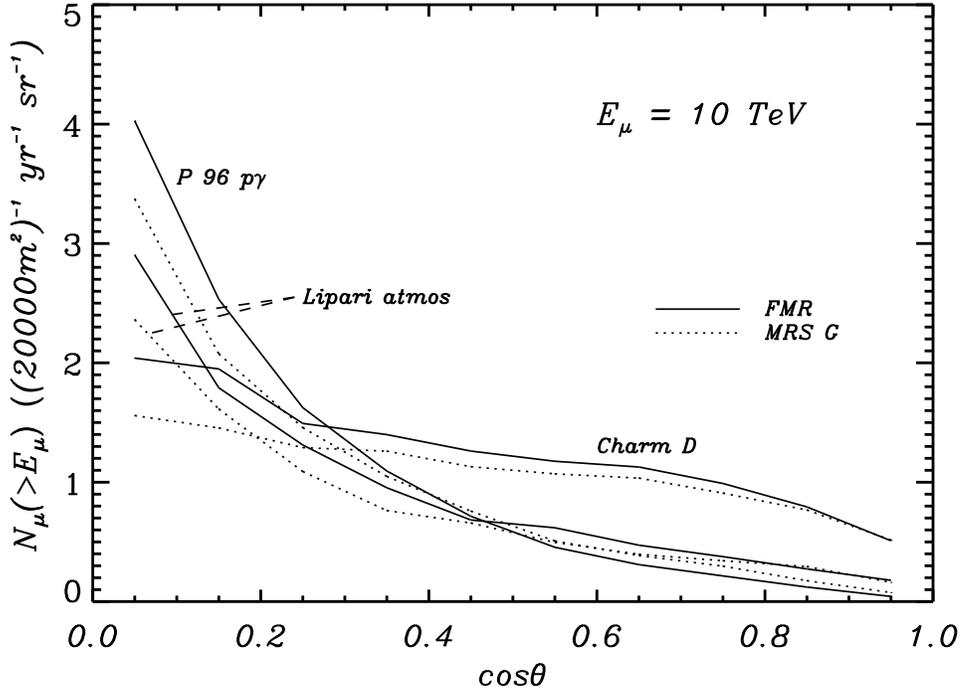}
\caption{\label{10TeVblow} Muon angular distributions for a
$20000$m$^2$ downward
looking detector employing a
 muon energy cut of $10^4$ GeV. The curve labelled {\em Charm D} is
an upper estimate of the muons from the prompt decay of charmed mesons into
neutrinos in the atmosphere and  could be up to two orders
of magnitude smaller.
}
\end{figure}

 The  P96$p\gamma$, Charm D and  conventional atmospheric fluxes
produce different angular
distributions, shown in figure \ref{10TeVblow}.
 The shape of the atmospheric neutrino
induced muon curve reflects primarily the angular distribution of the
source neutrinos. The isotropic charm-induced neutrinos produce a flatter
angular distribution; but with muons resulting from higher energy neutrinos
the angular shape shows evidence of
 some neutrino absorption. The muons induced by the 
isotropic P96$p\gamma$ flux show a strong angular dependence due to 
contributions from much higher neutrino energies. The three
 angular distributions
are quite different for nadir angles greater than $60^o$ ($\cos\theta < 0.5$).
A detector located deep enough such that atmospheric muons are not a problem
should be sensitive to these differences, provided enough muons were
measured to reveal the angular shape (requiring many years exposure for
a 20000 m$^2$ detector or a much larger detector). 

Surface or shallow detectors
are affected by misreconstruction of downward muons, which restricts their
acceptance aperture. Figure \ref{10TeVblow}
 shows that for nadir angles less than
$60^o$ ($\cos\theta > 0.5$) the angular distributions of all three spectra
are the same shape (fairly flat).  Thus, even using a very large detector/long
exposure time but with restricted aperture,  the angular distribution alone 
is not sufficient
 to 
determine  how much of a small
excess of muons  was due to AGN, because of  the large 
uncertainty in the  charm induced neutrino flux. 
 
Measuring  the energy spectrum of the observed muons may allow
a differentiation between source spectra.
The P96$p\gamma$ and Charm D fluxes give equal numbers of muons above
10 TeV; however $\sim$95\% of the Charm D muons are in the energy range
10-100 TeV. The P96$p\gamma$ muons are split roughly equally between the 
range 10-100 TeV and the region above 100 TeV. A comparison of the excesses
in each of these energy regions may aid identification. However, we are 
dealing with very low count rates where statistical fluctuations and the
uncertainty in the conventional atmospheric neutrino flux will also play 
a role.  

As we have just seen, determining
 the origin of a small excess of muons above the conventional
atmospheric expectation relies on looking for angular or energy 
distribution characteristics. The P96$p\gamma$ and Charm D spectra produce
 sufficiently different muon distributions to allow them to be 
distinguished.
The integral muon energy spectrum (figure \ref{intmuonrates})
for the P96$p\gamma/pp$ flux 
is similar to
the Charm D muon spectrum, although it produces an absolute level of
muons about a factor of 10 bigger. Also the angular
distributions  for the P96$p\gamma/pp$ flux have near the same slopes as the
Charm D angular distributions.
 This shows that the higher the 
contribution from $pp$ interactions to the diffuse AGN flux, the more
the muon energy and angular distributions become similar to those from the
charm induced neutrinos. An AGN spectrum made up of only 10\% $pp$ and 
90\% $p\gamma$ contributions produces a differential muon spectrum with
$\sim$85\% of the muons (above 10 TeV) in the energy range 10-100 TeV, and an absolute 
level only a few times bigger than the Charm D muon spectrum. 
 This  shows that a small observed muon excess showing angular and energy
distribution characteristics typical of the charm induced muon spectra 
could actually be due to a combination of an AGN spectrum with a modest $pp$ 
contribution with a low overall flux level (of order of the P96$p\gamma$
flux given here) and a charm spectrum lower than the Charm D spectrum 
considered here.

In this discussion I have  considered models which produce 
low numbers of muons in large detectors. Clearly, there will be
unambiguous signatures of AGN fluxes if the more optimistic 
fluxes are correct. Excesses well above those predicted by the 
Charm D flux will be clear evidence of an extra terrestrial origin.
The P96$p\gamma$ flux, although of the same level
as the Charm D flux, has distinctive characteristics 
(flatter muon energy spectrum and steeper angular distribution) 
that set it apart from the charm induced muons.
 However, AGN spectra with only modest $pp$ contributions produce muon 
distributions with similar characteristics to the charm induced muons.
  Interpreting a small observed excess of muons above the conventional
atmospheric expectation will require a better understanding of the
production of neutrinos from the prompt decay of charmed particles 
in the 
atmosphere.


\section{Conclusions}
In this work I have calculated the expected AGN neutrino induced muon rates
in large detectors and have compared them to the backgrounds expected
from atmospheric neutrinos (produced by the decay of pions, kaons, muons 
and the prompt decay of charmed mesons).
 A Monte Carlo simulation has been used which correctly accounts for the
effects of neutral current scattering and uses cross sections calculated
with the FMR and MRS G parton distributions, which were derived from 
recent accelerator data.
 These higher cross sections result in greater muon rates due to a
greater probability of muon production in the final kilometres before
a detector, but for incident
neutrino directions in the 120 degree aperture about the nadir this effect
is largely offset by increased neutrino absorption. The increases in muon rates
come from nadir angles greater than 60 degrees, the region where surface
or shallow detectors are most susceptible to atmospheric muon contamination.
The very deepest detectors will be able to observe these increased fluxes.

The lowest neutrino flux considered, due to $p\gamma$ interactions only in 
the Protheroe AGN model, would produce roughly  6 muons per year above 
10 TeV in a
20000 m$^2$ detector, which is about the same as that expected
 from the conventional atmospheric 
sources and the same as expected from
the highest estimates from charmed meson decay. 
The lowest neutrino induced
muon flux
due to charmed meson decay  gives practically negligible muon rates, which
suggests that telling these sources apart may be difficult.
 However, the P96$p\gamma$ AGN 
flux produces muons with angular
 and energy characteristics that would allow it 
to be distinguished from the charm and conventional atmospheric 
backgrounds. On the other hand,
 an AGN flux with around 10\% or more contribution
from $pp$ interactions produces muon angular and energy  distributions
similar to those of the charm induced muons. An observation of muons 
seemingly consistent with the level of the Charm D flux could actually be
due to an AGN spectrum with a modest $pp$ contribution and an overall 
flux level around 1/10 th of the P96$p\gamma/pp$ flux, in combination with
a charm induced flux consistent with the TIG model. A better understanding
of the charm production mechanism in the atmosphere will then be required to 
interpret such an observation of a small number of muons above the 
conventional atmospheric background.

 This discussion has centred around a possible 
\lq worst case' scenario (low
AGN fluxes and uncertain charm fluxes).
The larger fluxes  considered in this work would be readily observable
in a 20000 m$^2$ detector, using energy cuts of around 1-10 TeV.
 The planned deployments or extensions of the 
DUMAND, AMANDA, NESTOR and
Lake Baikal experiments should herald the birth of observational
extra-terrestrial neutrino astronomy, if the more optimistic
 AGN models and neutrino
fluxes
considered in this work are near to correct.

\section{Acknowledgments}   
I    thank 
F. Steffens, W. Melnitchouk, A. Thomas and G. Frichter for useful discussions
on QCD and neutrino cross sections and R. Protheroe for clarifying aspects
of AGN and atmospheric neutrino production.
I thank B. Dawson and R. Clay  for useful discussions,
 encouragement and careful
reading of manuscripts and M. Roberts,
   T. Porter and J. Reid for their advice
 throughout the course of this work.
 An anonymous referee is also thanked for helpful comments on the
original manuscript.


\begin{thebibliography}{10}

\bibitem{SDSS91}
F.W. Stecker, C.~Done, M.H. Salamon, and P.~Sommers.
\newblock {\em Phys. Rev. Lett.} 66 (1991) 2697,
\newblock  Errata {\em Phys. Rev. Lett.} 69 (1992) 2738 

\bibitem{SDSSH92}
F.W. Stecker, C.~Done, M.H. Salamon, and P.~Sommers.
\newblock  {\em
  Proceedings of the {W}orkshop on {H}igh {E}nergy {N}eutrino {A}stronomy},
  page~1. University of Hawaii, World Scientific, March 1992.

\bibitem{SS95}
F.W. Stecker and  M.H. Salamon.
\newblock  {\em astro-ph/9501064} 
{\em Space Science Reviews} (submitted)  (1995)


\bibitem{SPH92}
A.P. Szabo and R.J. Protheroe.
\newblock  {\em
  Proceedings of the {W}orkshop on {H}igh {E}nergy {N}eutrino {A}stronomy},
  page~24. University of Hawaii, World Scientific, March 1992.

\bibitem{SP94}
A.P. Szabo and R.J. Protheroe.
\newblock {\em Astropart. Phys.} 2 (1994) 365


\bibitem{Roberts92}
A.~Roberts.
\newblock {\em Rev. Mod. Phys.} 64 (1992) 259

\bibitem{AMANDA95}
P.C. Mock et~al.
\newblock In {\em Proceedings of the 24th {I}nternational {C}osmic {R}ay
  {C}onference}, Rome, {I}taly, 1 (1995) 758

\bibitem{Baikal95}
I.A. Belolaptikov et~al.
\newblock In {\em Proceedings of the 24th {I}nternational {C}osmic {R}ay
  {C}onference}, Rome, {I}taly, 1 (1995) 742

\bibitem{NESTOR95}
S.~Bottai et~al.
\newblock In {\em Proceedings of the 24th {I}nternational {C}osmic {R}ay
  {C}onference}, Rome, {I}taly, 1 (1995) 1080



\bibitem{StanevH92}
T.~Stanev.
\newblock {\em
  Proceedings of the {W}orkshop on {H}igh {E}nergy {N}eutrino {A}stronomy},
  page 354. University of Hawaii, World Scientific, March 1992.

\bibitem{Stanev94}
T.~Stanev.
\newblock {\em Nucl. Phys. B (Proc. Suppl.)} 35 (1994) 185

\bibitem{Stenger91}
V.J. Stenger.
\newblock In {\em Proceedings of the {W}orkshop: {T}rends in {A}stroparticle
  {P}hysics}, Aachen, Germany, October 1991.

\bibitem{ZHV92}
E.~Zas, F.~Halzen and R.A. V\'{a}zquez
\newblock {\em Astropart. Phys. } 1 (1993) 297




\bibitem{GHS95}
T.K. Gaisser, F.Halzen, and T.Stanev.
\newblock {\em Phys. Rep.} 258 (1995) 173

\bibitem{FMRradio}
G.M. Frichter, D.W. McKay, and J.P. Ralston.
\newblock {\em Phys. Rev. D} 53  (1996) 1684


\bibitem{GQRS95}
R.~Gandhi  et~al.
\newblock {\em hep-ph/9512364} {\em Astropart. Phys.} (submitted) (1995)

 

\bibitem{P96}
R.J. Protheroe.
\newblock In{\em Proceedings of the IAU Colloquium 163,
Accretion Phenomena and Related Outflows,} edited by
D. Wickramasinghe et al,   1996, in press


\bibitem{MRSG}
A.D. Martin, W.J. Stirling, and R.G. Roberts.
\newblock {\em Phys. Lett. B} 354 (1995) 155

\bibitem{FMR95}
G.M. Frichter, D.W. McKay, and J.P. Ralston.
\newblock {\em Phys. Rev. Lett.} 74 (1995) 1508

\bibitem{H192}
I.~Abt et~al ({H}1~{C}ollaboration)
\newblock {\em Nucl. Phys. B} 407(1993) 515

\bibitem{ZEUS92}
M.~Derrick et~al. ({ZEUS C}ollaboration)
\newblock {\em Phys. Lett. B} 316 (1993) 412

\bibitem{H193}
T.~Ahmed et~al ({H1} {C}ollaboration)
\newblock {\em Nucl. Phys. B} 439 (1995) 471

\bibitem{ZEUS93}
M.~Derrick et~al ({ZEUS C}ollaboration)
\newblock {\em Z. Physik C - Part. and Fields} 65 (1995) 379


\bibitem{Lipari93}
P.~Lipari.
\newblock {\em Astropart. Phys.} 1 (1993) 195




\bibitem{TIG95}
M.~Thunman, G.~Ingelman and P.~ Gondolo.
\newblock{\em Nucl. Phys. B Proc. Suppl. } 43 (1995) 274

\bibitem{CTEQ}
H.~Lai et~al.
\newblock {\em Phys. Rev. D} 51 (1995) 4763




\bibitem{TDLee}
T.D. Lee.
\newblock {\em Particle Physics and Introduction to Field Theory}.
\newblock Harwood Academic, Switzerland, 1981.

\bibitem{AH89}
I.J.R. Aitchison and A.J. Hey.
\newblock {\em Gauge Theories in Particle Physics: a Practical Introduction}.
\newblock Adam Hilger, Bristol and Philadelphia, 2 edition, 1989.

\bibitem{CG69}
C.G. Callan and D.J. Gross.
\newblock {\em Phys. Rev. Lett.} 22 (1969) 156

\bibitem{GG87}
T.K. Gaisser and A.F. Grillo.
\newblock {\em Phys. Rev. D} 36 (1987) 2752

\bibitem{PD94}
Particle~Data Group.
\newblock Review of {P}article {P}roperties.
\newblock {\em Phys. Rev. D} 50 (1994)

\bibitem{deGroot79}
J.G.H. de~Groot et~al.
\newblock {\em Z. Physik C - Part. and Fields} 1 (1979) 143

\bibitem{Buras80}
A.J. Buras.
\newblock {\em Rev. Mod. Phys.} 52 (1980) 199


\bibitem{RQ88}
M.H. Reno and C.~Quigg.
\newblock {\em Phys. Rev. D} 37 (1988) 657

\bibitem{QRW86}
C.~Quigg, M.H. Reno, and T.P. Walker.
\newblock {\em Phys. Rev. Lett.} 57 (1986) 774

\bibitem{RenoH92}
M.H. Reno.
\newblock  {\em
  Proceedings of the {W}orkshop on {H}igh {E}nergy {N}eutrino {A}stronomy},
  page 211. University of Hawaii, World Scientific, March 1992.

\bibitem{GL72}
V.N. Gribov and L.N. Lipatov.
\newblock {\em Sov. J. Nucl. Phys.} 15 (1972) 675
\newblock {\em Yad. Fiz.} 15 (1972) 1218

\bibitem{GL72b}
V.N. Gribov and L.N. Lipatov.
\newblock {\em Sov. J. Nucl. Phys.} 15 (1972) 438
\newblock {\em Yad. Fiz.} 15 (1972) 781

\bibitem{AP77}
G.~Altarelli and G.~Parisi.
\newblock {\em Nucl. Phys. B} 126 (1977) 298

\bibitem{Dok77}
Y.~L Dokshitser.
\newblock {\em Sov. Phys. JETP} 46 (1977) 641
\newblock {\em Zh. Eksp. Teor. Fiz.} 73 (1977) 1216

\bibitem{MRS93}
A.D. Martin, W.J. Stirling, and R.G. Roberts.
\newblock {\em Phys. Rev. D} 47 (1993) 867

\bibitem{MRS93b}
A.D. Martin, W.J. Stirling, and R.G. Roberts.
\newblock {\em Phys. Lett. B} 306 (1993) 145

\bibitem{MRS94}
A.D. Martin, W.J.Stirling, and R.G.Roberts.
\newblock {\em Phys. Rev. D} 50 (1994) 6734


\bibitem{FKL81}
E.G. Floratos, C.~Kounnas, and R~Lacaze.
\newblock {\em Nucl. Phys. B} 192 (1981) 417

\bibitem{Gluck90}
M.~Gl\"{u}ck, E.~Reya, and A.~Vogt.
\newblock {\em Z. Physik C - Part. and Fields} 48 (1990) 471

\bibitem{GLR83}
L.V. Gribov, E.M. Levin, and M.G. Ryskin.
\newblock {\em Phys. Rep.} 100 (1983) 1

\bibitem{Wally95}
T.~Weigel and W.~Melnitchouk.
\newblock Submitted to {\em Nucl. Phys. B} (1995)

\bibitem{SS83}
M.~Samorski and W.~Stamm.
\newblock {\em Ap. J.} 268 (1983) L17

\bibitem{JLE83}
J.~Lloyd-Evans et~al.
\newblock {\em Nature} 305 (1983) 784

\bibitem{PCG84}
R.J. Protheroe, R.W. Clay, and P.R. Gerhardy.
\newblock {\em Ap. J.} 280 (1984) L47

\bibitem{Yodh92}
G.B. Yodh.
\newblock  {\em
  Proceedings of the {W}orkshop on {H}igh {E}nergy {N}eutrino {A}stronomy},
  page 257. University of Hawaii, World Scientific, March 1992.

\bibitem{GS85}
T.K. Gaisser and T.~Stanev.
\newblock {\em Phys. Rev. Lett.} 54 (1985) 2265

\bibitem{GS85a}
T.K. Gaisser and T.~Stanev.
\newblock {\em Phys. Rev. D} 31 (1985) 2770

\bibitem{KTW85}
E.W. Kolb, M.S. Turner, and T.P. Walker.
\newblock {\em Phys. Rev. D} 32 (1985) 1145

\bibitem{CASA92}
J.W. Cronin et~al.
\newblock {\em Phys. Rev. D} 45 (1992)

\bibitem{CASA95a}
A.~Borione et~al.
\newblock In {\em Proceedings of the 24th {I}nternational {C}osmic {R}ay
  {C}onference},  Rome, {I}taly, 2 (1995) 439

\bibitem{CASA95b}
A.~Borione et~al.
\newblock In {\em Proceedings of the 24th {I}nternational {C}osmic {R}ay
  {C}onference},  Rome, {I}taly, 2 (1995) 503

\bibitem{CASA95c}
A.~Borione et~al.
\newblock In {\em Proceedings of the 24th {I}nternational {C}osmic {R}ay
  {C}onference},  Rome, {I}taly, 2 (1995) 430

\bibitem{CASA95d}
A.~Borione et~al.
\newblock In {\em Proceedings of the 24th {I}nternational {C}osmic {R}ay
  {C}onference},  Rome, {I}taly, 2 (1995) 435

\bibitem{MD91}
T.~Maccacaro et~al.
\newblock {\em Ap. J.} 374 (1991) 117

\bibitem{MT89}
K.~Morisawa and F.~Takahara.
\newblock {\em Pub. Astron. Soc. Japan} 41 (1989) 873

\bibitem{EGRET}
J.~Chiang et~al.
\newblock{\em Ap. J.} 452 (1995) 156

\bibitem{SAS2}
D.J. Thompson and C.E. Fichtel.
\newblock {\em Astron. Astroph. } 109 (1982) 352

\bibitem{KMRS}
J.~Kwiecinski et~al.
\newblock {\em Phys. Rev. D} 42 (1990) 3645

\bibitem{BGZR86}
V.S. Berezinsky, A.Z. Gazizov, G.T. Zatsepin and I.L. Rozental.
\newblock {\em Sov. J. Nucl. Phys.} 43 (1986) 406

\bibitem{MyPhD}
G.C. Hill.
\newblock{\em PhD thesis, University of Adelaide} (1996)

\bibitem{PREM}
A.M. Dziewonski and D.L. Anderson.
\newblock {\em Phys.  Earth  Planet. Inter.} 25 (1981) 297


\bibitem{SL91}
P.~Lipari and T.~Stanev.
\newblock {\em Phys. Rev D} 44 (1991) 3543

\bibitem{PS68}
A.A. Petrukhin and V.V. Shestakov.
\newblock {\em Can. J. Phys.} 46 (1968) S377
\newblock (10th {ICRC} Calgary).

\bibitem{KP71}
R.P. Kokoulin and A.A. Petrukhin.
\newblock {\em Proceedings of the 12th {I}nternational {C}osmic {R}ay
  {C}onference ({H}obart)}, 6  (1971) 2436

\bibitem{Bezru81}
L.B. Bezrukov and \'{E}.V. Bugaev.
\newblock {\em Yad. Fiz.} 33 (1981) 1195
\newblock Sov. J. Nucl. Phys. 33 (1981)

\bibitem{Stenger92}
V.J. Stenger.
\newblock {\em Proceedings of the 1992 {NESTOR} {W}orkshop, {P}ylos, {G}reece},
  1992.

\bibitem{Rhode94}
W.~Rhode et~al.
\newblock {\em Astropart. Phys.} (submitted) (1995)

\bibitem{Rhode95}
W.~Rhode et~al.
\newblock In {\em Proceedings of the 24th {I}nternational {C}osmic {R}ay
  {C}onference}, Rome, {I}taly, 1 (1995) 781


\bibitem{Gaisser93}
T.K. Gaisser.
\newblock {\em Nucl. Phys. B (Proc. Suppl.)} 31 (1993) 399


\end{thebibliography}
\end{document}